\documentclass{article}

\usepackage{arxiv}

\usepackage[utf8]{inputenc} 
\usepackage[T1]{fontenc}    
\usepackage{hyperref}       
\usepackage{url}            
\usepackage{booktabs}       
\usepackage{amsfonts}       
\usepackage{nicefrac}       
\usepackage{microtype}      
\usepackage{lipsum}		
\usepackage{graphicx}
\usepackage[round,sort,numbers]{natbib}
\usepackage{doi}

\usepackage{graphicx}
\usepackage{algorithmic}
\usepackage[ruled,vlined]{algorithm2e}
\SetKwInput{KwInput}{Input}                
\SetKwInput{KwOutput}{Output}              
\usepackage{amssymb}
\usepackage{amsmath}
\usepackage{subfigure}

\title{Co-Embedding: Discovering Communities on Bipartite Graphs through Projection}

\date{} 					

\author{
	\href{https://orcid.org/0000-0002-9072-1535}{\includegraphics[scale=0.06]{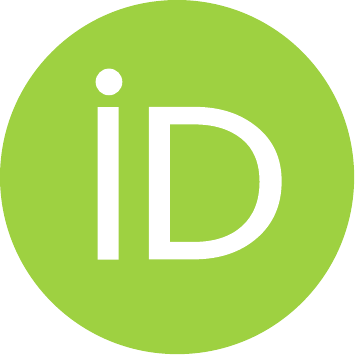}\hspace{1mm}Ga\"elle Candel} \\
	Wordline TSS Labs, Paris \\
	\texttt{firstname.lastname@worldline.com} \\
	\& \\
	D\'epartement d'informatique de l'ENS \\
	ENS, CNRS, PSL University, Paris \\
	\texttt{firstname.lastname@ens.fr}\\
	\And
  David Naccache \\
	D\'epartement d'informatique de l'ENS \\
	ENS, CNRS, PSL University, Paris \\
	\texttt{firstname.lastname@ens.fr}\\
}

\renewcommand{\headeright}{}
\renewcommand{\undertitle}{}
\renewcommand{\shorttitle}{Co-Embedding: Discovering Communities through Projection}

\hypersetup{
pdftitle={Co-Embedding: Discovering Communities through Projection},
pdfsubject={cs.AI, cs.ML},
pdfauthor={Gaëlle Candel, David Naccache},
pdfkeywords={bipartite graphs, co-clustering, co-embedding, dimension relatedness, visualization},
}

\begin{document}
\maketitle

\begin{abstract}
	Many datasets take the form of a bipartite graph where two types of nodes are connected by relationships, like the movies watched by a user or the tags associated with a file.
	The partitioning of the bipartite graph could be used to fasten recommender systems, or reduce the information retrieval system's index size, by identifying groups of items with similar properties.

	This type of graph is often processed by algorithms using the Vector Space Model representation, where a binary vector represents an item with $0$ and $1$.
	The main problem with this representation is the dimension relatedness, like words’ synonymity, which is not considered.

	This article proposes a co-clustering algorithm using items projection, allowing the measurement of features similarity.
	We evaluated our algorithm on a cluster retrieval task.
	Over various datasets, our algorithm produced well balanced clusters with coherent items in, leading to high retrieval scores on this task.
\end{abstract}


\section{Introduction}

Many datasets can be represented as a bipartite  graph (BPG), making the links between two different types of items.
This could be the movies a user watched, the purchases he made, the music he listened at.
Outside of user interaction, it could be tags associated with a file, such as the keywords of an article, or the genes with which a molecule interacts.

\textit{Community discovery} or graph partitioning helps improving scalability in different contexts.
Collaborative Filtering Recommender Systems (CF RecSys) work by suggesting new items to a user based on the similarity of its history to the other users' history.
Without optimization, a user is compared to all the other users to find those most similar to him.
Clustering reduces costs by looking at similar users within the same partition rather than looking at them in the entire dataset \cite{Sarwar2002RecommenderSF}.
Clustering is not limited to users and has similar benefits when clustering items \cite{Conner1999ClusteringIF}, or both together \cite{George2005ASC}.
These approaches are based on the clustering hypothesis \cite{Vdorhees1985TheCH} supposing that similar documents would answer the same information needs.
Cluster-based retrieval systems are based on this same hypothesis \cite{Chen2005CLUECR,Liu2004ClusterbasedRU}.
A very large database that cannot be stored in a single server could be split into multiple servers,  hosting a particular thematic cluster.
A query would be compared to an index with clusters' summary and routed to the most relevant server, reducing the overall number of operations.

There are different approaches to cluster a bipartite graph.
The algorithms can be classified into two distinct categories: \textit{one-way} and \textit{two-way} partitioning algorithms, with the latter grouping both sides of the graph simultaneously.
In the \textit{one-way} approach, the set of nodes belonging to the same type are clustered, without considering the clusters that would be obtained when clustering the nodes from the other type.
The nodes to cluster are often represented into a matrix using the Vector Space Model (VSM) \cite{Salton1975AVS}, where the nodes of the second type are considered as their features \cite{Ravindran2015KMeansDC} or used to construct the features \cite{Lei2019PatentAB}.
The resulting matrix can be exploited to cluster items with usual algorithms such as $k$-means, measure the items' proximity using cosine similarity measure, or project items using dimensionality reduction algorithms \cite{Lei2019PatentAB}.

Another possibility is to convert the $2$-mode graph into a $1$-mode graph by collapsing the graph.
The conversion to a unipartite graph is convenient as many classical community detection algorithms can be used \cite{Sariyce2018PeelingBN}.
The nodes from a single type are preserved and new weighted edges are inferred from the bipartite graph.
There are multiple approaches to convert the bipartite graph into a unipartite graph \cite{Latapy2008BasicNF,Arroyo2020GraphMB}.
However, the main problem is the growing number of edges and the loss of information due to the collapse \cite{Sariyce2018PeelingBN,Latapy2008BasicNF}.

In contrast, \textit{two-way} clustering algorithms -- also called co-clustering or simultaneous clustering \cite{Charrad2011SimultaneousCA} -- exploit the raw structure to cluster both types of nodes simultaneously.
\textit{Two-way} approaches lead to better results, even if the goal is to partition only one side of the BPG \cite{Charrad2011SimultaneousCA,Dhillon03InformationCoclustering} by exploiting synergies between the obtained clusters.

Among co-clustering approaches, the majority of the literature focuses on \textit{bi-clustering} approaches, where a bicluster is characterized by a particular set of rows and columns when the BPG is represented under the matrix form.
Therefore, a bicluster gathers items of heterogeneous type.
There exist sub-categories characterized by the possibility to overlap clusters and to assign an item to multiple clusters  \cite{Kaiser2011BiclusteringMS}.
The majority of the algorithms creates biclusters with exclusive rows and columns \cite{Zha2001BipartiteGP,Nie2017LearningAS,Sariyce2018PeelingBN}.
In this situation, the models assume a one-to-one match between sample and feature clusters and strong connectivity, forming blocks when visualizing the matrix with items grouped by clusters.
This clustering type is -- to some extent -- equivalent to finding sub-graphs forming dense structures \cite{Sariyce2018PeelingBN} that could exist without the other groups.

In contrast, the literature covering the case where a cluster gathers items of the same type is limited.
While this case seems to correspond to the one-way partitioning, it highly differs from it as a row (column) cluster is expressed as a mixture of column (row) clusters.
In contrast, the one-way approach considers features individually.
Among this group, we can mention non-negative matrix factorization (NMF) approaches \cite{Chen2010NonNegativeMF,Du2017DCNMFNM}.
This approach offers a more flexible framework than the latent block model leading to a checkerboard representation when re-ordering rows and columns by groups.
Nonetheless, this approach assumes that there are latent variables connecting rows to columns; therefore we obtain the same number of row and column clusters.

In this work, we particularly focus on the case where the number of row and column clusters are not necessarily equal.
These approaches are more flexible as the number of clusters is not constrained by the reciprocal type of items and increase the number of possible configurations.
In this group, we can mention \cite{Dhillon01co-clusteringdocuments} combining a spectral decomposition to a $k$-means, and \cite{Dhillon03InformationCoclustering} based on information theory minimizing the loss of mutual information between clustered and unclustered items.
These models require as input the number of clusters to find, limiting their usability without prior knowledge.
The work of \cite{SELOSSE2020106866} extends the latent block model to cluster heterogeneous data types (ordinal, continuous, count).
Samples are clustered into groups without specific constraints, while features are clustered with features of the same type.
The number of feature clusters is automatically inferred and independent from the number of sample clusters, leading to a flexible automatic co-clustering approach.

Most of these approaches are designed for very general cases and do not assume the underlying nature of BPG.
More specifically, they do not consider \textit{dimension relatedness}.
If animals are features like [tiger, lion, frog], the distance between their respective binary vectors $[1, 0, 0]$, $[0, 1, 0]$ and $[0, 0, 1]$ is $2$ for all possible pairs despite the similarity between \textit{tiger} and \textit{lion}.
The different features are considered as orthogonal, while it is far from the reality where some characteristics are often correlated, even weakly.

\textit{Dimension relatedness} has been addressed mainly for textual applications.
The works of \cite{Gabrilovich2007ComputingSR,Tsatsaronis2009AGV} propose using an external database (WordNet and Wikipedia resp.) to measure dimension relatedness.
Over a very large corpus, the dimension correlation could be learned, as proposed in the work of \cite{abbasi2009richvsm}.
However, an external database is not always available, depending on the language used,  and the type of objects considered.
The work of \cite{Sidorov2014SoftSA} suggest learning the similarities between features by directly measuring their textual similarity.
Feature similarity is derived from the Levenshtein distance between the feature names.

These different textual approaches learn the distances between features, but none between samples.
One would like to apply the same treatment to samples and features in a co-clustering approach exploiting the sample-feature duality.

In this article, we propose a co-clustering algorithm that addresses the problem of dimension relatedness.
The proposed approach follows a co-embedding process, where each side of the bipartite graph is projected in a low dimensional space.
This projection enables to measure items relatedness based on their features' location.
The process alternates between projecting each side of the graph, until invariance compared to the previous embedding.

The rest of this article is organized as follows.
First, the algorithm is detailed in the following section,
then the datasets and evaluation methods are presented.
Visual, numerical, and textual results are then presented in the experimental section,
followed by possible extensions in the discussion and a conclusion.

\section{Co-Embedding}

\paragraph{Notations:}
A bipartite graph $\mathcal{G} = (V^I, V^{II}, E)$ is a graph composed of two types of nodes $V^I = \{v^I_i\}_{i=1:|V^I|}$ and $V^{II} = \{v^{II}_i\}_{i=1:|V^{II}|}$.
The edges $E$ connecting nodes only exist between nodes of different types $E \subseteq V^{I} \times V^{II}$.
The existence of a link between two nodes $v^I \in V^I$ and $v^{II} \in V^{II}$ is denoted as $\delta(v^I, v^{II}) = 1 \text{ if } (v^I, v^{II}) \in E \text{ else } 0$.
The representation of $\mathcal{G}$ under the VSM model is $M = [\delta(v^I, v^{II})]_{v^I \in V^I, v^{II} \in V^{II}}$
The number of occurrences of $v^I \in V^I$ is denoted $|v^I| = \sum_{v^{II}_i \in V^{II}} \delta(v^I, v^{II}_i)$ and defined similarly for a node of $V^{II}$.

\paragraph{Sample-Feature Duality:}

A \textit{feature} is relative to the \textit{sample}'s type considered.
Nodes of $V^I$ are features of $V^{II}$ and $V^{II}$ of $V^I$.
Therefore, the equations would be written in terms of samples $\mathcal{S}$ and features $\mathcal{F}$ instead of $V^I$ and $V^{II}$, as our process inverses periodically the roles.
As most of the datasets used correspond to (tag, resources) pairs, we would refer to \textit{tags} $\mathcal{T}$ and \textit{resources} $\mathcal{R}$ when a difference needs to be underlined.

\paragraph{Proposed Approach:}
We address the problem of dimension relatedness for co-clustering a bipartite graph
by projecting the items into a low dimensional space.
The samples relatedness is measured by comparing their features' location on the embedding space.
Based on their similarity, samples are next embedded into another low dimensional space using the $t$-SNE algorithm \cite{vanDerMaaten2008} which is a key point in the process.
Next, samples and features exchange their role to start a new iteration.
The process is repeated several times until the neighborhood around each item is stable.
The last step of our method is an automated clustering using the Mean-Shift algorithm \cite{Mean_shift} over the two last co-embedding.

The details of each step will be detailed in this section.
First, we start with the $t$-SNE algorithm to provide an understanding of the embedding properties.
Next, we detail how features' relatedness is measured and how it is used to measure samples similarity.
Last, we explain the procedure for Mean-Shift clustering concluding the co-clustering process.

\subsection{$t$-SNE Embedding}

These paragraphs describe the general ideas behind the $t$-SNE algorithm \cite{vanDerMaaten2008}.
For a set of $n$ items $X$, this non-parametric embedding algorithm transforms it into a low dimensional representation $Y \in \mathbb{R}^{n \times d}$, with  $d$ often set to $2$ for visualization purposes:
\begin{equation} \label{eq:tsne}
  Y \leftarrow t\text{-SNE}(D(X); d, perp)
\end{equation}
The \textit{perplexity} parameter ($perp$)  controls the number of nearest neighbors in $X$ to preserve in $Y$.
This prevents highly connected items of the graph from having too many neighbors and improves the neighborhood of weakly connected items.
The algorithm works by minimizing the Kullback-Leibler divergence between the image matrices of $X$ and $Y$, obtained respectively using a Gaussian kernel and a $t$-Student kernel.
This kernel asymmetry creates repulsive long-range forces leading to well-separated groups.
Another characteristic is the homogeneous scaling over an embedding, allowing to measure similarity the same way independently of the location and the crowdedness.

The perplexity governs the embedding shape, where a large value focuses on large scale structures, while a lower one on details.
The understanding of \textit{large} is relative to the number of items, and could be adapted to tags and resources with $perp^{\mathcal{R}}$ and $perp^{\mathcal{T}}$ as they do not necessarily have the same size.

As this algorithm tries to preserve local neighborhood relationships, nothing can be said on long-range distances.
However, the proximity between items in the embedding can be exploited as an alternative to distances in the high dimensional space.

\subsection{Representing Samples using their Features Embedding}

The binary vector of a sample $s$ is transformed into a vector of probabilities.
The feature embedding $Y^{\mathcal{F}} = \{\mathbf{y}_i\}_{i=1:|\mathcal{F}|}$ is used to create this vector by taking into account three factors:
the feature's location, the feature's popularity, and the related features.
The vector representing $s$ is $\mathbf{p}(s) = [p(f | s)]_{f \in \mathcal{F}}$, where the term concerning a feature $f$ is defined as:
\begin{equation}\label{eq:proba}
  p(f | s) = \frac{1}{c(s)} \sum_{f_i \in \mathcal{F}} \frac{\delta(s, f_i)}{|f_i|} K(\mathbf{y}, \mathbf{y}_i)
\end{equation}
where $\frac{\delta(s, f_i)}{|f_i|}$ represents the contribution of feature $f_i$ contained in $s$.
The normalization constant $c(s)$ ensures that $\sum_{f \in \mathcal{F}} p(f | s) = 1$.
For a feature $f_i$ of $s$ with image $\mathbf{y}_i$, the kernel redistributes the feature's mass on the neighborhood feature $f$ with image $\mathbf{y}$ based on their kernelized distance $K(\mathbf{y}, \mathbf{y}_j)$.
Therefore, the weight $p(f | s)$ can be non-zero even if $f$ is not a feature of $s$.
\textit{Mass redistribution} assumes that all existing edges are \textit{true} edges and some edges are missing, i.e. there are no misconnected items, but the information available is incomplete.
The kernel allows us to consider the unlinked items with some degree of confidence based on their proximity.

\paragraph{Kernel Choice}

The $t$-SNE embedding uses a $t$-Student kernel to map items.
While it seems a natural kernel choice, this distribution has a long tail, allowing distant items to contribute.
In the $t$-SNE algorithm, this kernel was chosen to create long-range forces for better clusters' separation.
As the goal is to identify closely related items, the use of a Gaussian kernel is more adapted, defined as $K(\mathbf{y}_i, \mathbf{y}_j; \sigma) = \exp\left(-\frac{\|\mathbf{y}_i - \mathbf{y}_j\|^2}{2 \sigma^2}\right)$ with bandwidth parameter $\sigma$.
This kernel has the advantage to be more localized and adaptable using $\sigma$.

\paragraph{$\pmb{\sigma}$'s Choice:}

The perplexity impacts the distances between items within an embedding.
Consequently, $\sigma$ is adapted by looking at the effective distances by:
\begin{equation} \label{eq:radius}
  \hat{\sigma}(Y, k) = \text{Median}\left[\|\mathbf{y} - \mathbf{y}_{k\text{-NN}}\|\right]
\end{equation}
where $\mathbf{y}_{k\text{-NN}}$ is the $k$-est nearest neighbors of point $\mathbf{y}$, with $k = \lfloor perp \rceil$ and $\|\mathbf{y}\| = \sqrt{\sum_{i=1}^d y_i^2}$ is the Euclidian norm.
The use of the median rather than the mean limits the outliers' contribution which would enlarge $\sigma$.

\subsection{Building the Samples Embedding}

The $t$-SNE algorithm requires as input a distance matrix.
Therefore, the distances between samples are obtained by measuring the divergence between the samples' probability vector obtained using Eq. \eqref{eq:proba}.
We use the Jeffrey-Kullback-Leibler divergence to measure items proximity, with the formulation:

\begin{equation} \label{eq:JKL}
  KL^J\left(\mathbf{p}(s_a) \| \mathbf{p}(s_b)\right) = \frac{1}{2}\left(KL(\mathbf{p}(s_a) \| \mathbf{p}(s_b)) + KL(\mathbf{p}(s_b) \| \mathbf{p}(s_a))\right)
\end{equation}

with $\mathbf{p}(s_a)$ and $\mathbf{p}(s_b)$ the vectors of sample $s_a$ and $s_b$ respectively.
We use this divergence instead of the traditional KL because of the symmetry of $KL^J$.
The matrix $D^S = [KL^J\left(\mathbf{p}(s_a) \| \mathbf{p}(s_b)\right)]_{s_a, s_b \in \mathcal{S}}$ allows to obtain a new sample embedding $Y^{\mathcal{S}} \leftarrow t\text{-SNE}(D^{\mathcal{S}})$  used in the next iteration inverting samples and features' role.

\subsection{Embedding Procedure Summary \& Parameter Choices}

\begin{algorithm}[H]
\DontPrintSemicolon

  \KwInput{$(perp^{\mathcal{R}}, perp^{\mathcal{T}})$: Perplexities; $k$: Number of iterations.
  }
  \KwOutput{$Y^{\mathcal{R}}, Y^{\mathcal{T}}$: Resources and tags respective embedding}
  \KwData{$M = \{\delta(r, t)\}_{r \in \mathcal{R}, t \in \mathcal{T}}$: the resources-tags matrix.}

  $Y^{\mathcal{R}}(0),Y^{\mathcal{T}}(0) \leftarrow \text{Init}(M)$ \tcp*[r]{Initialization}
  $(\mathcal{S}, \mathcal{F}) \leftarrow (\mathcal{R}, \mathcal{T})$ \tcp*[r]{Resources start the role of samples}

  \For{$i=1$ \KwTo $2k$}
  {
    $\sigma^{\mathcal{F}} \leftarrow f(Y^{\mathcal{F}}, perp^{\mathcal{F}})$ \tcp*[r]{Using Eq. \eqref{eq:radius}}
    $P^{\mathcal{S}} = \{\mathbf{p}(s)\}_{s=1:|\mathcal{S}|} \leftarrow g(M, Y^{\mathcal{F}}, \sigma^{\mathcal{F}})$ \tcp*[r]{Using Eq.  \eqref{eq:proba}}
    $D^{\mathcal{S}} \leftarrow [KL^J(\mathbf{p}_a \| \mathbf{p}_b)]_{\mathbf{p}_a, \mathbf{p}_b \in P^{\mathcal{S}}}$ \tcp*[r]{Using Eq. \eqref{eq:JKL}}
    $Y^{\mathcal{S}} \leftarrow t\text{-SNE}(D^{\mathcal{S}}; perp^{\mathcal{S}})$
    \\
    \tcc{Exchange roles}
    $(\mathcal{S}, \mathcal{F}) \leftarrow (\mathcal{F}, \mathcal{S})$
    $M \leftarrow M^T$
  }
  $(Y^{\mathcal{R}}, Y^{\mathcal{T}}) \leftarrow (Y^{\mathcal{S}}, Y^{\mathcal{F}})$
\caption{Co-Embedding procedure} \label{pseudocode}
\end{algorithm}

\paragraph{Initialization:}

At the start, no initial embedding exists yet, which prevents from measuring density.
We initialize the first embedding with the $d$ first eigen-vectors of $M$ obtained by SVD.
Compared to a random initialization, it fastens the convergence process by starting from an organized state. In pseudocode \eqref{pseudocode}, resources start the role of samples.
This is an arbitrary choice and has almost no impact on the final result.

\paragraph*{Iteration}

At each step, the samples' probability density is estimated over the feature's space.
Then, samples' divergence matrix is obtained using these probabilities.
The embedding step finish by embedding using the $t$-SNE algorithm to obtain a new sample embedding $Y^{\mathcal{S}}(t+1)$ using the divergence matrix as input.
After building the embedding, features and samples exchange their respective roles.

\paragraph{Ending Criterion:}

The pseudocode \eqref{pseudocode} iterates $k$ times on each type of item, an arbitrary value around a dozen steps, as the algorithm does not minimize a particular criterion.
Nonetheless, the process can be monitored in terms of  \textit{neighborhood stability}, looking at if the neighborhood around a point is unchanged over successive embedding.
By denoting $\mathcal{N}_n(Y, i)$ the set of the $n$ nearest neighbors of $i$ in $Y$, we compare its neighborhood in $Y(t)$ and $Y(t+1)$ using the Jaccard similarity by $\frac{|\mathcal{N}_n(Y(t), i) \cap \mathcal{N}_n(Y(t+1), i)|}{|\mathcal{N}_n(Y(t), i) \cup \mathcal{N}_n(Y(t+1), i)|}$.

\paragraph{Algorithm Complexity:}

For $n$ samples and $m$ features, the probability estimation requires $\mathcal{O}(nm^2)$ operations.
The divergence measurement requires $\mathcal{O}(n^2m)$ operation to compute $n^2$ pairs with vectors of size $m$.
Then, the $t$-SNE embedding complexity is in $\mathcal{O}(kn^2)$ for $k$ update steps, typically between $100$ and $1000$, depending on the dataset size and convergence speed.
In total, a step described in pseudocode \ref{pseudocode} requires $\mathcal{O}\left(nm (n+m) + kn^2\right)$ operations.

\subsection{Clustering Embeddings}

The proposed algorithm leads to clusters' apparition when community structures exist in a dataset.
Groups are extracted using the Mean-Shift (MS) clustering algorithm \cite{Mean_shift}.
This algorithm follows an iterative process, moving items towards their \textit{mode}'s location, which are positions with maximal probability density.
The image $\bar{\mathbf{y}}$ of point $\mathbf{y}$ is moved toward its mode following the equation:
\begin{equation} \label{eq:mean_shift}
    \bar{\mathbf{y}} \leftarrow \frac{\sum_{j=1}^n K(\bar{\mathbf{y}}, \mathbf{y}_j) \mathbf{y}_j}{\sum_{j=1}^n K(\bar{\mathbf{y}}, \mathbf{y}_j)}
\end{equation}
starting with $\bar{\mathbf{y}} = \mathbf{y}$ and using the Gaussian kernel $K(.)$ defined previously, using the bandwidth parameter $\sigma$ estimated using Eq. \eqref{eq:radius}.
After several iteration steps, all items may have converged in some specific locations.
A cluster is obtained by gathering all items within a radius $\epsilon$.
Using MS, tags clusters $\mathcal{C}^T$ and resources clusters $\mathcal{C}^R$ are extracted from their respective last embedding $Y^T$ and $Y^R$.

\paragraph{Two-Ways Clustering:}
Mean-Shift is a \textit{one-way} clustering algorithm as samples and features are clustered independently.
Nevertheless, we consider the full process as a \textit{two-way} co-clustering algorithm as the embeddings are linked together,.
The MS algorithm has the advantage of being parameter-free, as the kernel bandwidth $\sigma$ is adapted to the embedding.
The two main advantages of MS is its ability to automatically discover the number of clusters and the uniqueness of the partitioning obtained.

\paragraph{Co-Clusters Relationships:}
A sample is represented by a mixture of features using Eq. \eqref{eq:proba}.
Two samples $s_a$ and $s_b$ are embedded close to each other area if they have a similar mixture.
The features of a sample are not necessarily all located in the same area.
Assuming the features of $s_a$ are located in $k$ distinct areas, $s_b$ is similar to $s_a$ only if its features are also located in these $k$ areas.
Therefore, we would obtain a co-cluster connected to $k$ features clusters.
If $s_b$ has no feature in one location or has features in another location out of this $k$ areas, its mixture would be highly dissimilar.
KL divergence highly penalizes couples of items where one has a mass where the other has none.
If $s_b$ has its features located in all these $k$ locations with many features in some of these locations, an asymmetry can appear due to the excess and deficit of mass in the different areas.
Depending on the strength of the asymmetry, $s_a$ and $s_b$ can be located either in the same cluster or in two distinct clusters.
A cluster can be characterized by the features clusters it is connected to and the strength of the connection.

\section{Evaluation Methods}

\subsection{Datasets}

We propose to evaluate our co-embedding approach over various datasets corresponding to tags associated with resources, as tags allow us to evaluate a group content subjectively.
We tried to select for each resource type two datasets to visualize the impact of  different data collection processes.
Some of the datasets are folksonomies, tagged by non-expert people with uncontrolled vocabulary, while the others are tagged by experts using a specific vocabulary.

Table \ref{tab:datasets} summarizes the different datasets' characteristics, such as the number of unique resources $|\mathcal{R}|$ and tags $|\mathcal{T}|$, the average number of tags per resources $\mathbb{E}(|r|)$ and the corresponding standard deviation $\sigma(|r|)$, and if the dataset is a folksonomy (\textit{Folks ?} column).

\begin{table}[hb]
  \caption{Summary of the (unprocessed) datasets' characteristics}
  \centering
  \label{tab:datasets}
  \begin{tabular}{|l||rr|rr|c|c|}
    \hline
    Dataset & $|\mathcal{R}|$ & $|\mathcal{T}|$ & $\mathbb{E}(|r|)$ & $\sigma(|r|)$  & Folks ? &  Media \\
    \hline
    BibTex    & 813,548   & 255,496  & 3.3 & 4.2   &  yes &  bibliography \\
    DBLP      & 4,894,081 & 132,337  & 10.2 & 1.7  &  no  &  bibliography \\

    Corel5K   & 5,000   & 347        & 3.5 & 0.6   &  no  &   image \\
    Flickr    & 946,113 & 345,897    & 11.5 & 9.6  &  yes &  image \\

    MovieLens & 10,381  & 1,127      & 44.3 & 27.7 &  no  &  movie \\
    IMDB      & 120,919 & 1,001      & 19.4 & 12.0 &  yes &  movie \\

    Delicious & 16,105  &   501      & 18.3 & 39.5 & yes &   URL \\
    BibUrl    & 618,245 & 156,497    &  3.4 &  2.7 & yes &   URL \\

    Last.fm   & 445,821 &  138,402   & 3.5 & 3.5   & yes &   music \\

    NG20      & 19,300  & 1,007      & 32.1 & 32.4 & no  &   netnews \\
    OHSUMED   & 13.929  & 1,002      & 39.7 & 17.1 & no  &   abstract \\
  \hline
\end{tabular}
\end{table}

For bibliographic tags, the \textit{BibTex} part of \textit{Bibsonomy} \cite{bibsonomy} and its non-folksonomy counterpart \textit{DBLP} \cite{DBLP} were selected.
For images, we used the  \textit{Flickr} folksonomy dataset (MIRFLICKR 25) and the \textit{Corel5K} \cite{corel5k}.
The \textit{IMDB} \cite{IMDB_cometa} and the \textit{MovieLens20M tag genome} \cite{movielens_20m} were used for movies.
\textit{IMDB} is classified as a folksonomy as words are extracted from movies' reviews, while \textit{MovieLens}'s tags come from a controlled list.
While $|\mathcal{T}|$ is similar for both, the vocabulary used are different.
Url datasets \textit{Delicious} \cite{delicious_cometa} and the \textit{URL} part of the \textit{Bibsonomy} database \cite{bibsonomy} (denoted \textit{BibUrl}) were selected.
We used the Million Song Dataset from \textit{Last.fm} \cite{million_song_dataset} for music songs but did not find another equivalent dataset.
We used textual datasets for comparison.
The \textit{NewsGroup20} \cite{newsgroup_cometa} (NG20) corresponds to news articles covering 20 different topics,
and the \textit{OHSUMED} \cite{ohsumed_cometa} to abstracts from MedLine papers.
The \textit{Corel5k}, \textit{Delicious}, \textit{IMDB}, \textit{NewsGroup20} and \textit{OHSUMED}  datasets were gathered on  \url{https://cometa.com}.

The detailed pre-processing steps are detailed in \ref{section:appendix}.
The largest datasets are sampled and filtered before use as they do not fit in memory.
The sampling and filtering steps are also detailed in the local appendix.

\subsection{Cluster Retrieval Tasks}

We propose to evaluate the co-clusters obtained using our method over a cluster-retrieval task,
where the goal is to find the cluster where the item's location is.
Additionally, we evaluate the ability of the features clusters to serve as a unit of description.
The sample's binary vector is replaced by the proportion of features it has in each cluster, providing a compact representation.
For a sample $s$, its description vector is:
\begin{equation} \label{eq:sample_compression}
  \mathbf{q}(s) = \left[\frac{\sum_{f \in C^F} \delta(s, f)}{\sum_{f \in F} \delta(s, f)}\right]_{C^F \in \mathcal{C}^F}
\end{equation}
as well as for a cluster $C \in \mathcal{C}^S$ by:
\begin{equation} \label{eq:group_compression}
  \mathbf{q}(C) = \left[\frac{\sum_{s \in C, f \in C^F} \delta(s, f)}{\sum_{s \in C, f \in F} \delta(s, f)}\right]_{C^F \in \mathcal{C}^F}
\end{equation}
The compressed vectors are denoted by $\mathbf{q}$ to avoid confusion with the vectors $\mathbf{p}$ defined in Eq. \eqref{eq:proba}.
We use the \textit{binary} features and not the \textit{diffused} ones here as the goal is to evaluate the partitioning obtained, and not the embedding quality.
Clusters $C \in \mathcal{C}^{\mathcal{S}}$ are sorted by increasing KL divergence $D(s, C) = KL(\mathbf{q}(s) \| \mathbf{q}(C))$.
The retrieval accuracy of this task is measured using the \textit{Mean Retrieval Rank} (MRR), which quickly drops with miss-prediction.

The retrieval task allows us to evaluate the partitioning relevance.
This is assessed by the capabilities of sample clusters to gather items with similar connections, and by the ability of feature clusters to serve as a description unit.
Cluster partitioning can help to speed up the retrieval process.
A user query is often short and inaccurate, which makes impossible an exact search.
The query must be compared to all $n$ items of the database to select the most relevant item.
By partitioning into $k$ clusters, the search requires $k$ operations to find the best cluster (assuming the correct cluster is well identified), and $\frac{n}{k}$ others to find the correct element within the cluster.
Therefore, the partitioning can help to reduce the search complexity from $\mathcal{O}(n)$ to $\mathcal{O}(k + \frac{n}{k})$, which could be beneficial for very large or distributed databases.

\subsection{Comparative Algorithm}

We compare our algorithm to the spectral co-clustering algorithm presented in \cite{Dhillon01co-clusteringdocuments}.
The algorithm uses the VSM representation that SVD  next decomposes.
Then, the singular vectors are clustered using $k$-Means.
The algorithm is adapted by grouping the two types of nodes separately.
As this algorithm has to enter the number of clusters to be searched, we use the number of clusters found using our approach.
$k$-Means is a non-deterministic algorithm leading to different results depending on the initialization seed.
The algorithm is run $10$ times, and the best result is kept for fairness.
We choose this algorithm as it has been widely used and has a strong theoretical basis but does not consider dimension relatedness.

\subsection{Evaluation Metrics}

\subsubsection{Cluster Quality}

Not all partitioning with $k$ clusters are equivalent.
The proposed task is very easy if a single cluster gathers all the items, and the other clusters are made of very small groups.
In contrast, the retrieval task using partitioning with clusters of equivalent size is much harder, as a random choice based on the mass would lead to very poor scores.
We define the partitioning entropy as the entropy relative to the clusters' size:

\begin{equation}\label{eq:coemb_entropy}
  H(\mathcal{C}) = - \sum_{C \in \mathcal{C}} Pr(C) \log_2\left(Pr(C) \right)
\end{equation}
where $Pr(C) = \frac{|C|}{\sum_{C_i \in \mathcal{C}} |C_i|}$, with values in $[0, \log_2 |\mathcal{C}|]$.

The value in terms of entropy is non-comparable to the number of clusters as it represents the number of bits necessary to encode the information.
The value is put to the power of $2$ to transform it into a meaningful value.
The effective number of cluster $k(\mathcal{C}) = 2^{H(\mathcal{P})} \in [1, |\mathcal{C}|]$, where a value of $1$ is synonymous with a large cluster, while a value close to $|\mathcal{C}|$ indicates that the clusters have a similar size.
This value better measures the retrieval difficulty, as clusters' size is considered.

\subsubsection{Items Representativeness}

The items in the identified clusters can be ranked by \textit{relevance} or \textit{representativeness}.
A sample is relevant if its distribution on the feature space is close to the mean distribution of the cluster to which it belongs to.

We propose to score items based on their KL-divergence from their respective cluster.
The KL-divergence could be expressed as $KL(A \| B) = H^*(A, B) - H(A)$, where $H^*(.)$ is the cross-entropy that represents the extra cost of coding $A$ with the optimal code of $B$. For a cluster $C\in \mathcal{C}^{\mathcal{S}}$, we define the representativeness of sample $s \in C$ relatively to $C$ as:
\begin{equation}\label{eq:representativeness}
  \text{repr}(s, C) = \max \left(0, 1 - \frac{KL(\mathbf{q}(s) \| \mathbf{q}(C))}{H(\mathbf{q}(s))} \right)
\end{equation}
A score $\text{repr}(s, C) \in [0, 1]$ is obtained, where $1$ represents the maximal relevance.

This normalization enables us to consider samples' frequency.
A low divergence between the item and the cluster is synonymous with high similarity.
However, underfrequent items are likely to have a lower divergence than frequent items.
By normalizing the divergence with the entropy, items are fairly compared to the cluster average.

\section{Experimental Results}

\subsection{Evolution over Time}

\begin{figure}[ht]
    \centering

    \subfigure[t=0]{\includegraphics[width=0.21\textwidth]{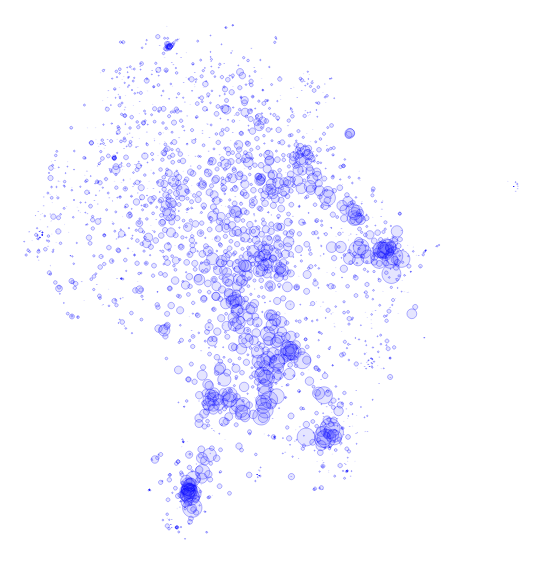}}
    \subfigure[t=2]{\includegraphics[width=0.21\textwidth]{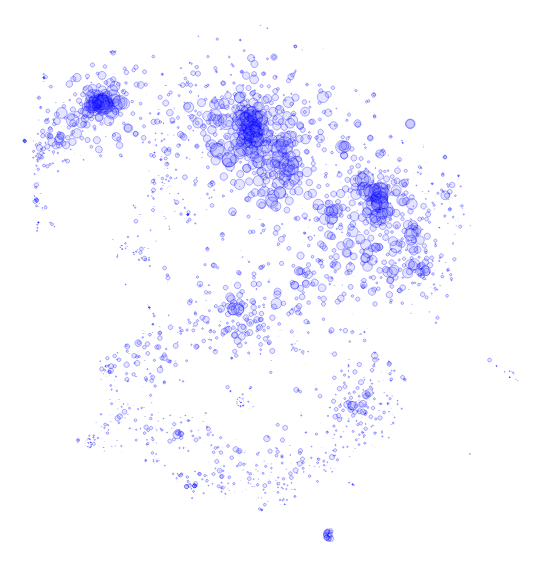}}
    \subfigure[t=4]{\includegraphics[width=0.21\textwidth]{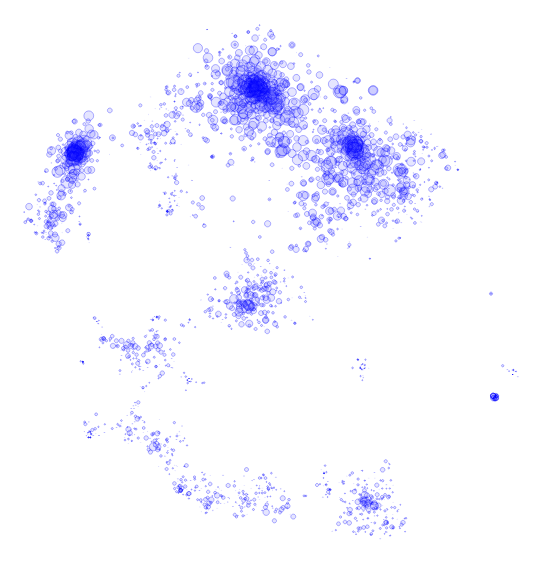}}
    \subfigure[t=6]{\includegraphics[width=0.21\textwidth]{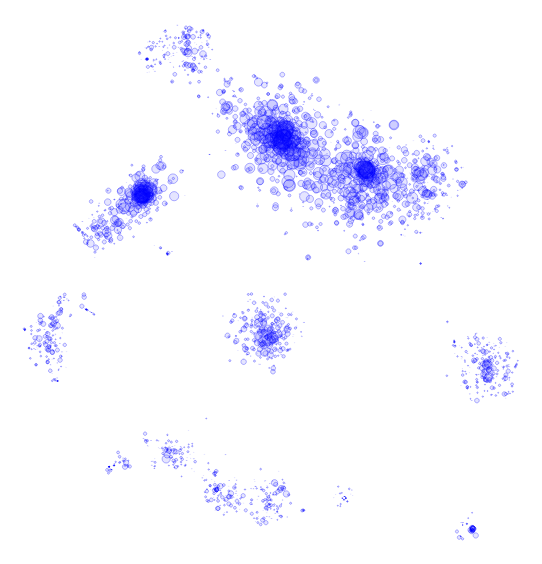}}

    \centering
    \subfigure[t=8]{\includegraphics[width=0.21\textwidth]{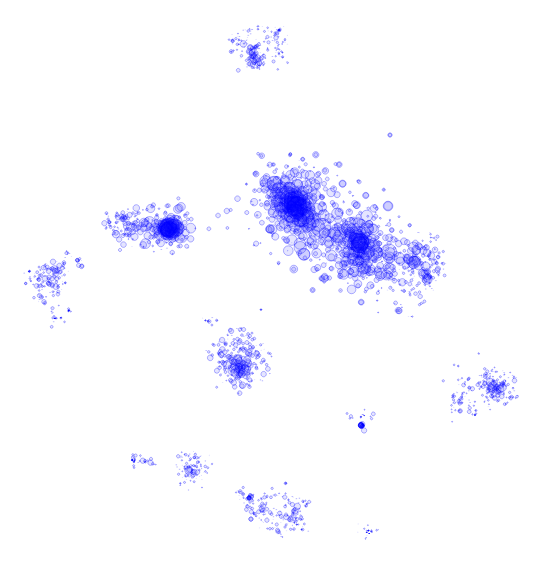}}
    \subfigure[t=10]{\includegraphics[width=0.21\textwidth]{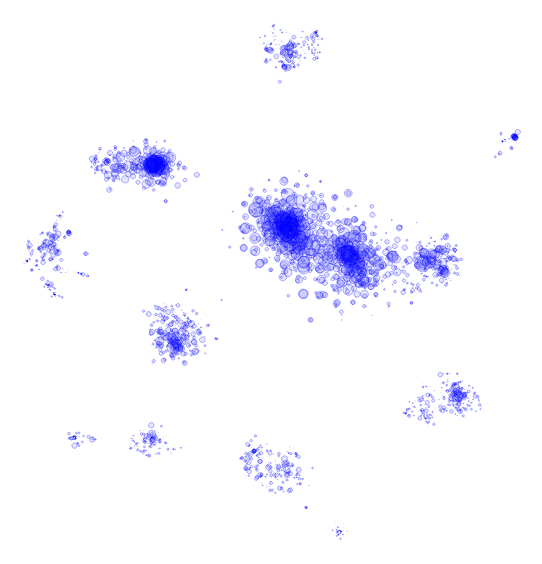}}
    \subfigure[t=12]{\includegraphics[width=0.21\textwidth]{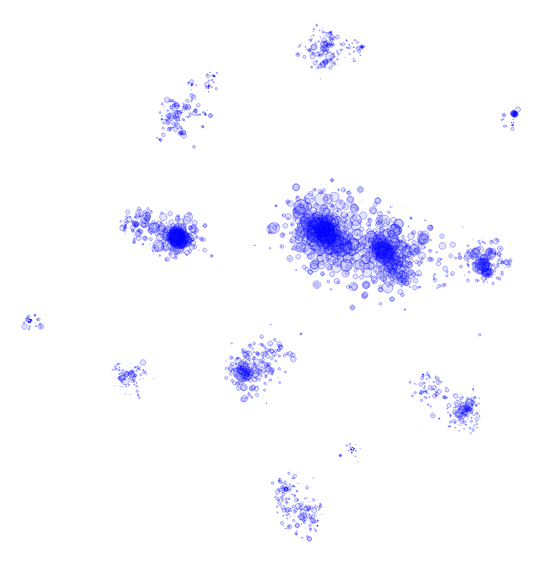}}
    \subfigure[t=14]{\includegraphics[width=0.21\textwidth]{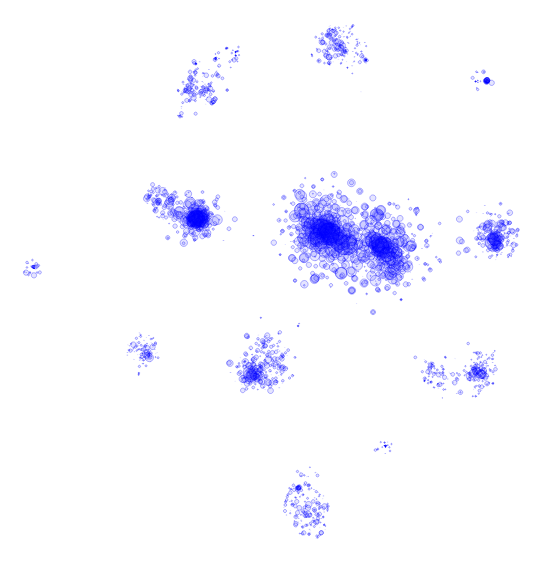}}

    \centering
    \caption{Evolution of the \textit{keywords} co-embedding for the \textit{Flickr} dataset.}
    \label{fig:coemb_timeline_keywords}
\end{figure}

\begin{figure}[ht]
    \centering
    \subfigure[t=1]{\includegraphics[width=0.21\textwidth]{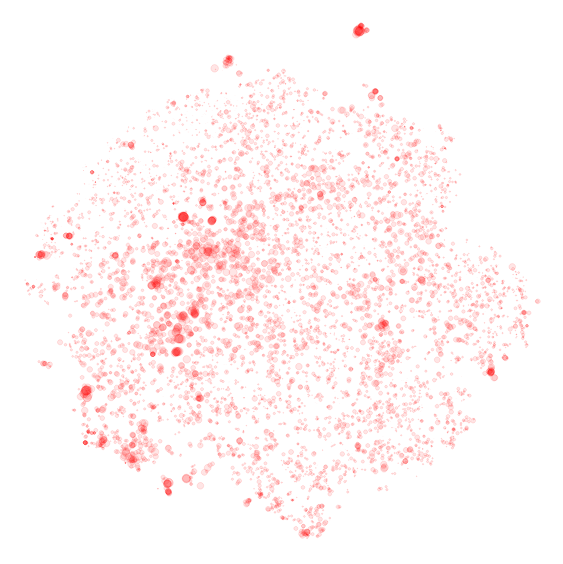}}
    \subfigure[t=3]{\includegraphics[width=0.21\textwidth]{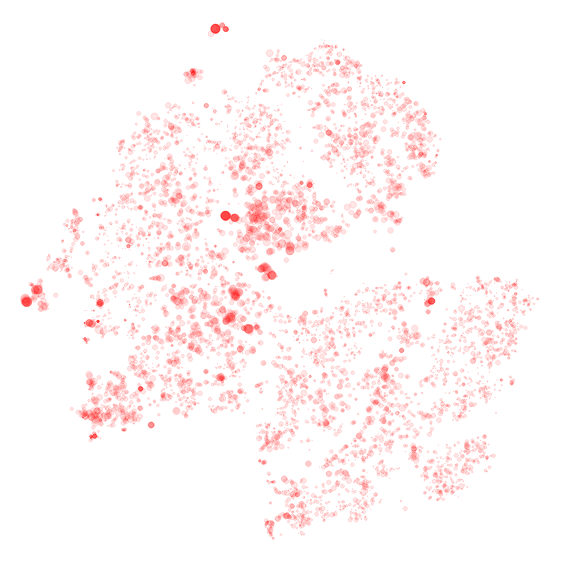}}
    \subfigure[t=5]{\includegraphics[width=0.21\textwidth]{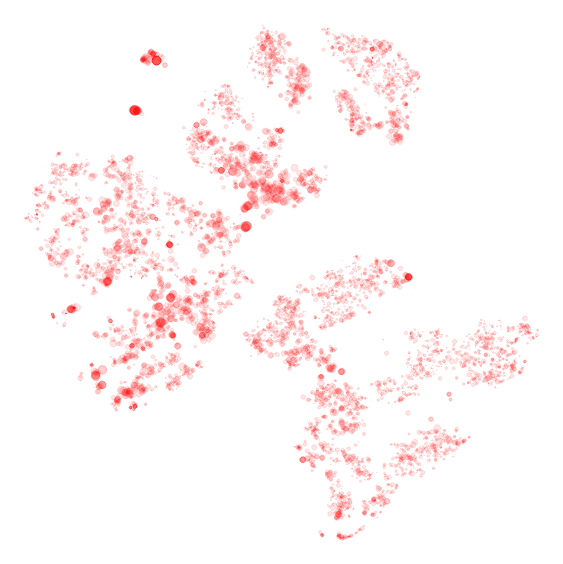}}
    \subfigure[t=7]{\includegraphics[width=0.21\textwidth]{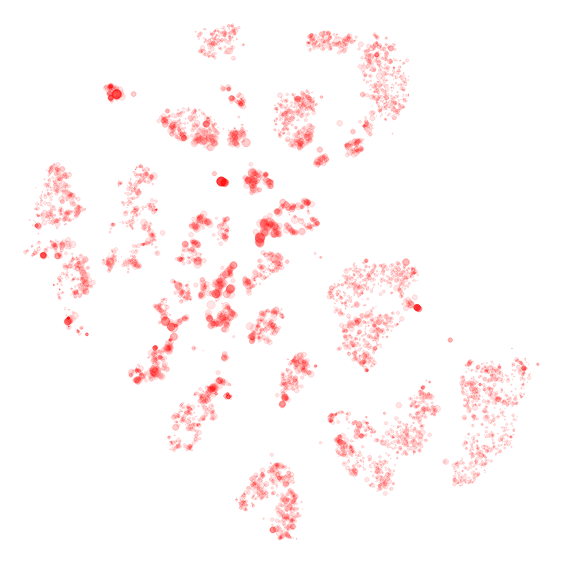}}

    \centering
    \subfigure[t=9]{\includegraphics[width=0.21\textwidth]{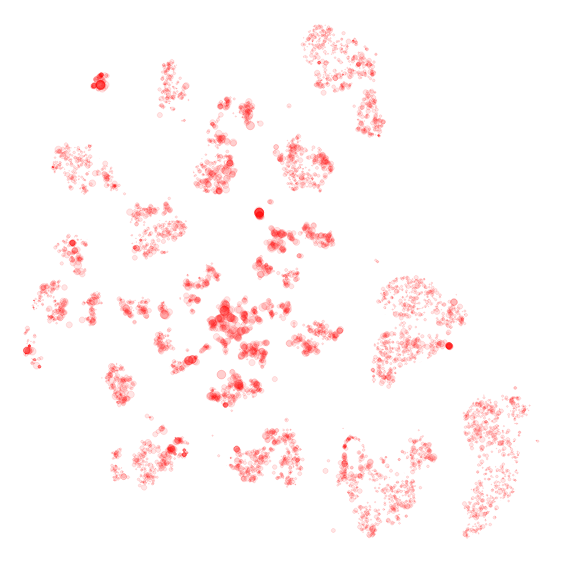}}
    \subfigure[t=11]{\includegraphics[width=0.21\textwidth]{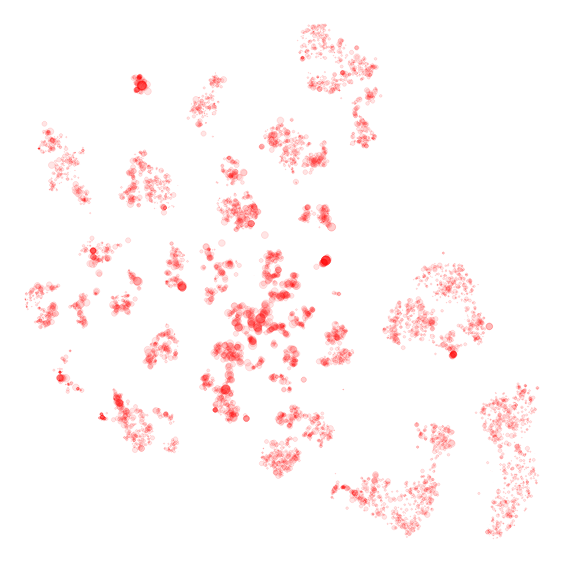}}
    \subfigure[t=13]{\includegraphics[width=0.21\textwidth]{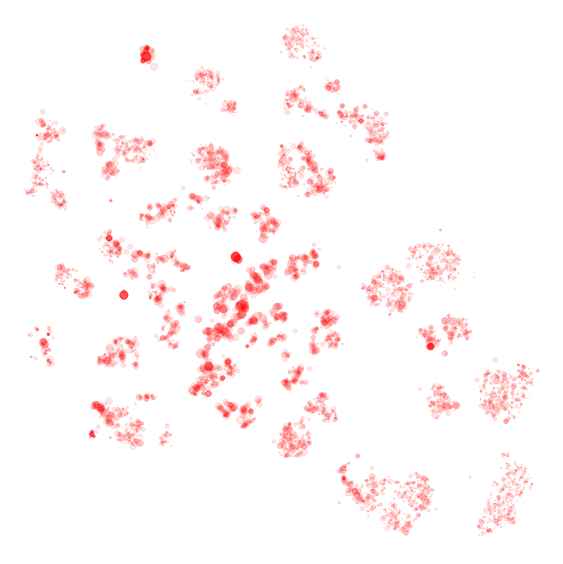}}
    \subfigure[t=15]{\includegraphics[width=0.21\textwidth]{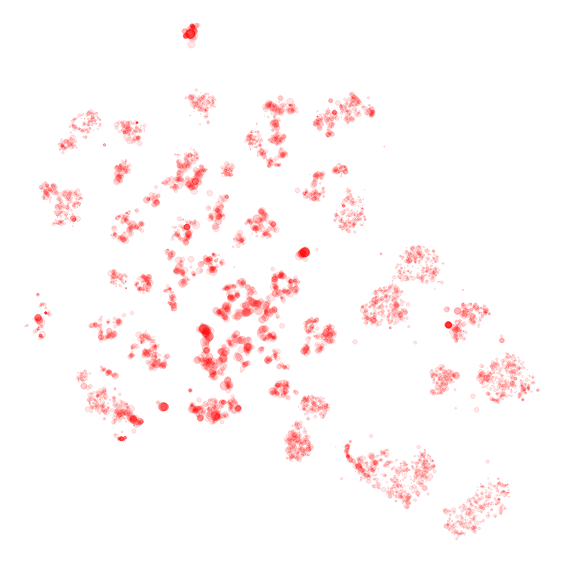}}

    \centering
    \caption{Evolution of the \textit{documents} co-embedding for the \textit{Flickr} dataset.}
    \label{fig:coemb_timeline_documents}
\end{figure}

Fig. \ref{fig:coemb_timeline_keywords} and Fig. \ref{fig:coemb_timeline_documents} represent the embedding chronology for keywords and resources respectively, for randomly selected items of the \textit{Flickr} dataset.
The subset is described in Table \ref{tab:CLS} and the resulting co-embedding is reused in the next experiments.

$t$-SNE is initialized with the previous items' location, allowing to visualize items displacement over time better.
Therefore, clusters' positions are approximately preserved over the successive steps.
The first keyword embedding ($t=0$) is obtained by exploiting the document eigen-vectors obtained by SVD decomposition, while the first document embedding ($t=1$) is obtained exploiting this first keyword embedding.

Major structures emerge from the initial mass of items during the first steps (until $t=6 \sim 7$).
During the last steps, clusters' shapes are refined.
The keyword clusters' locations are stable over time, but clusters tend to densify from step $t=6$ to step $t=14$, leading to well-separated clusters.
Keyword clusters' move from their previous location and some split into several pieces between $t=7$ and $t=15$.
At the end of the process, both embeddings show clusterable structure but differ in their spatial organization. The difference will be detailed in the next paragraphs.

\clearpage
\pagebreak

\subsection{Visual Results}

\begin{figure}
    \centering
    \subfigure[Corel5K]{\includegraphics[width=0.19\textwidth]{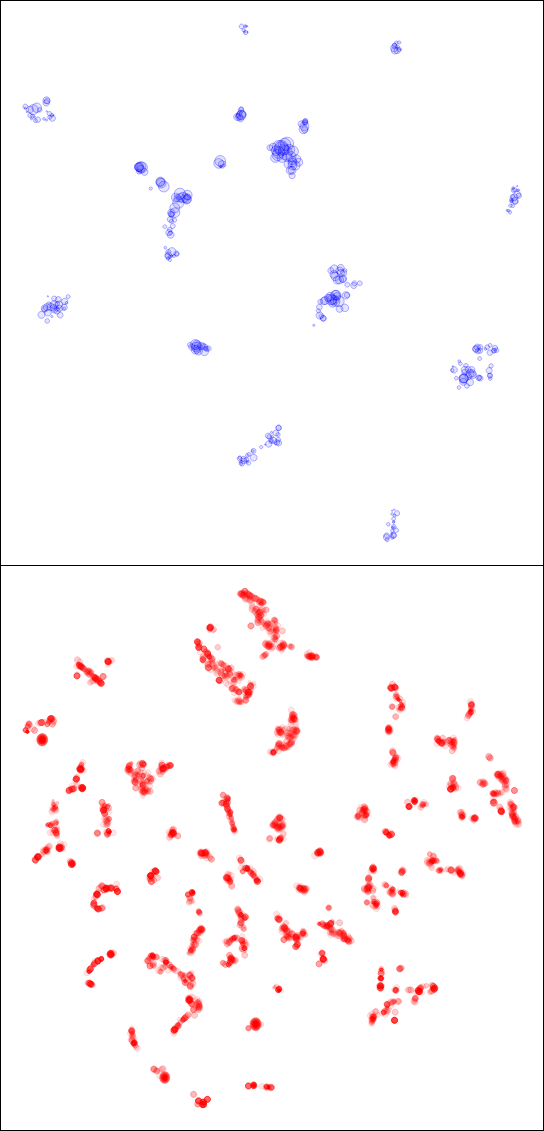}}
    \subfigure[Flickr]{\includegraphics[width=0.19\textwidth]{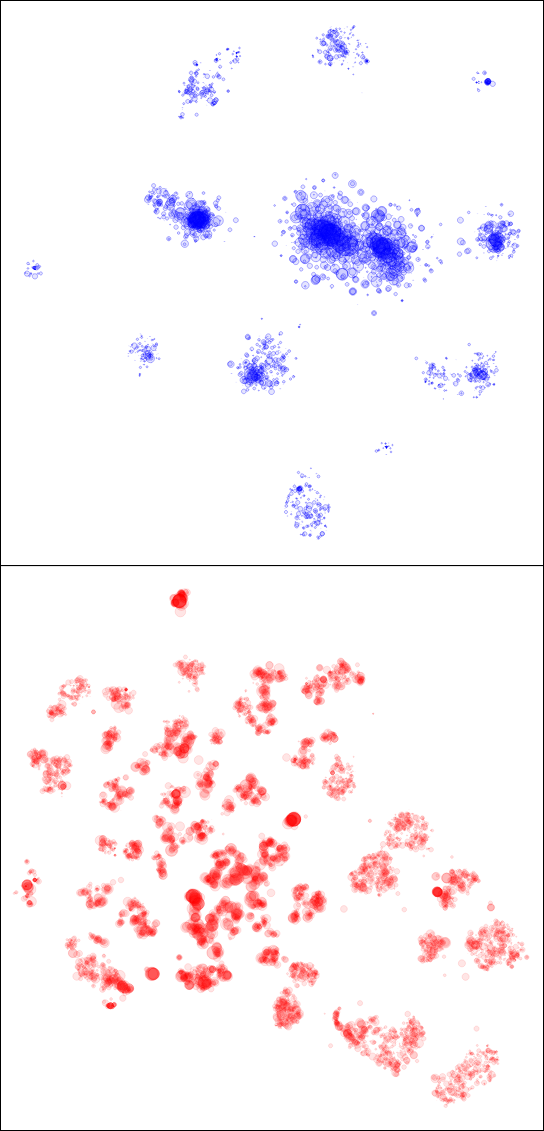}}
    \subfigure[IMDB]{\includegraphics[width=0.19\textwidth]{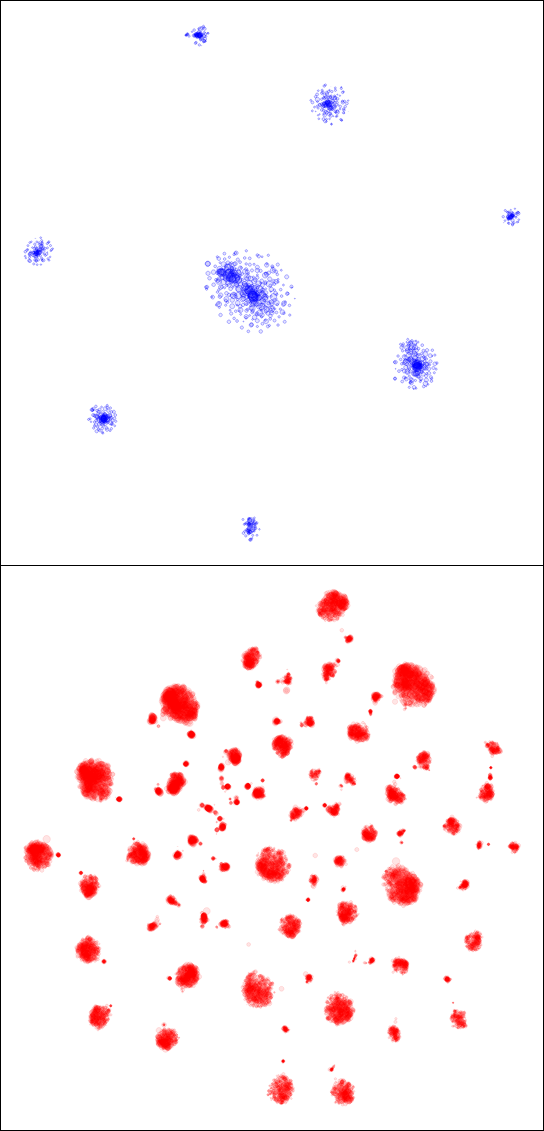}}
    \subfigure[MovieLens]{\includegraphics[width=0.19\textwidth]{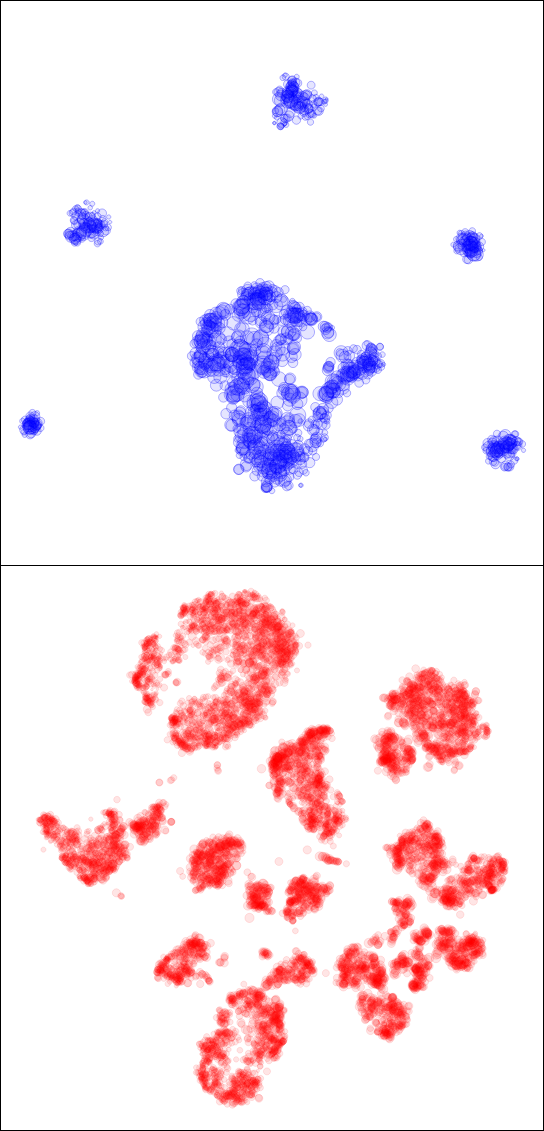}}

    \centering
    \subfigure[NG20]{\includegraphics[width=0.19\textwidth]{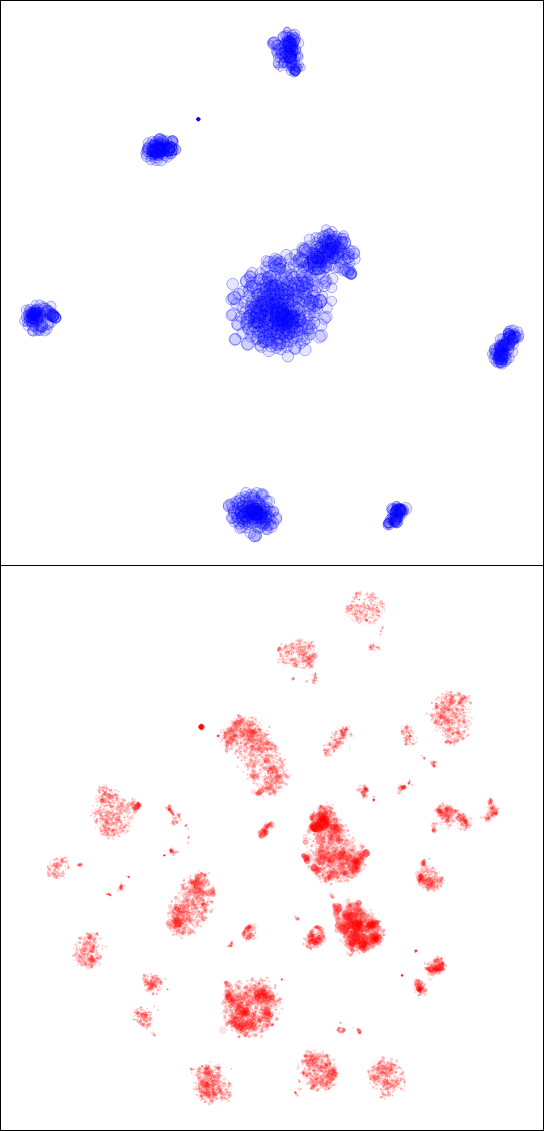}}
    \subfigure[OHSUMED]{\includegraphics[width=0.19\textwidth]{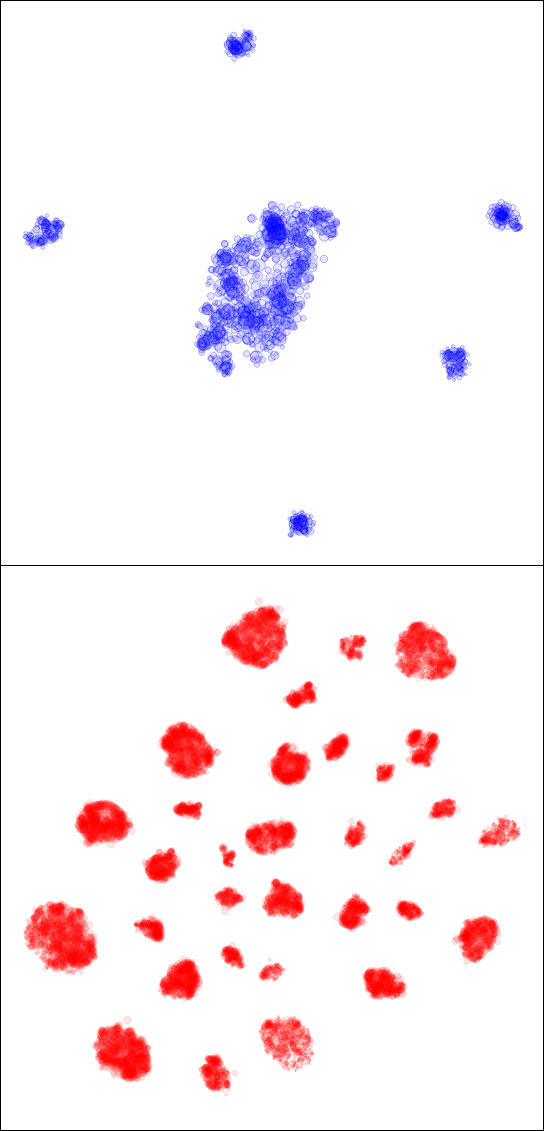}}
    \subfigure[DBLP (u)]{\includegraphics[width=0.19\textwidth]{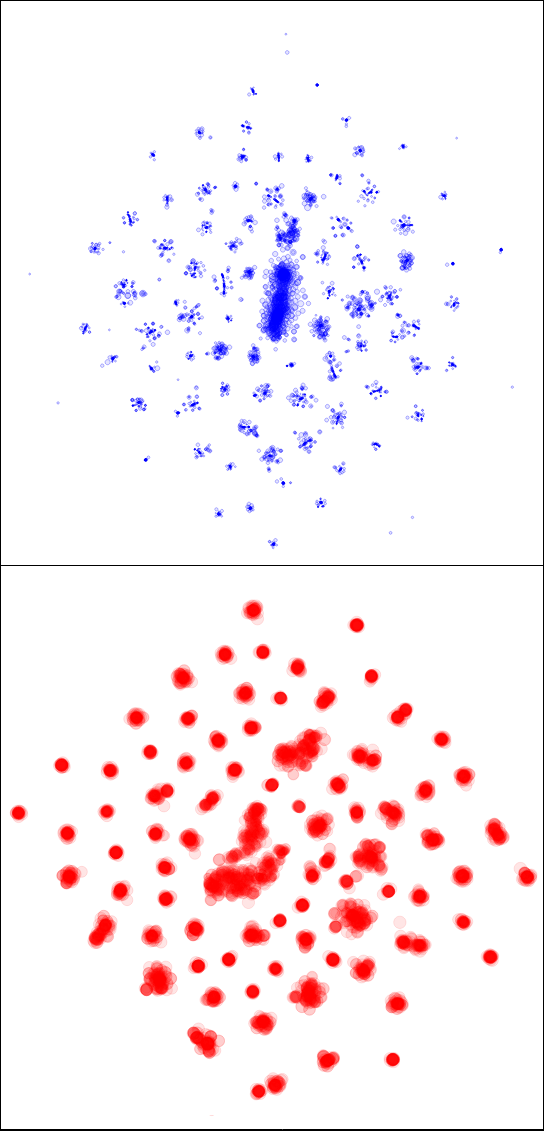}}
    \subfigure[DBLP (f)]{\includegraphics[width=0.19\textwidth]{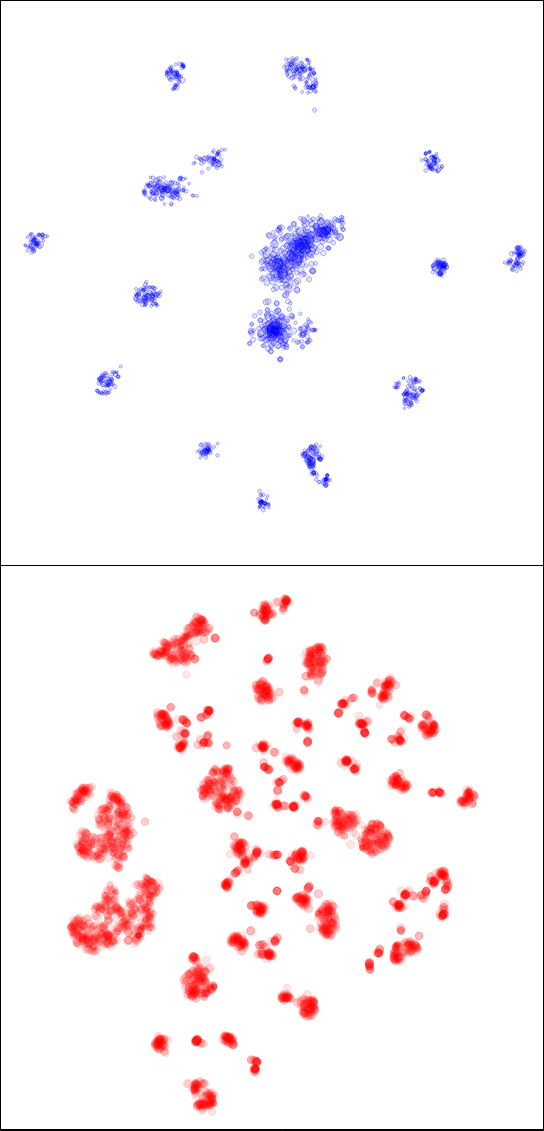}}

    \centering
    \subfigure[Delicious]{\includegraphics[width=0.19\textwidth]{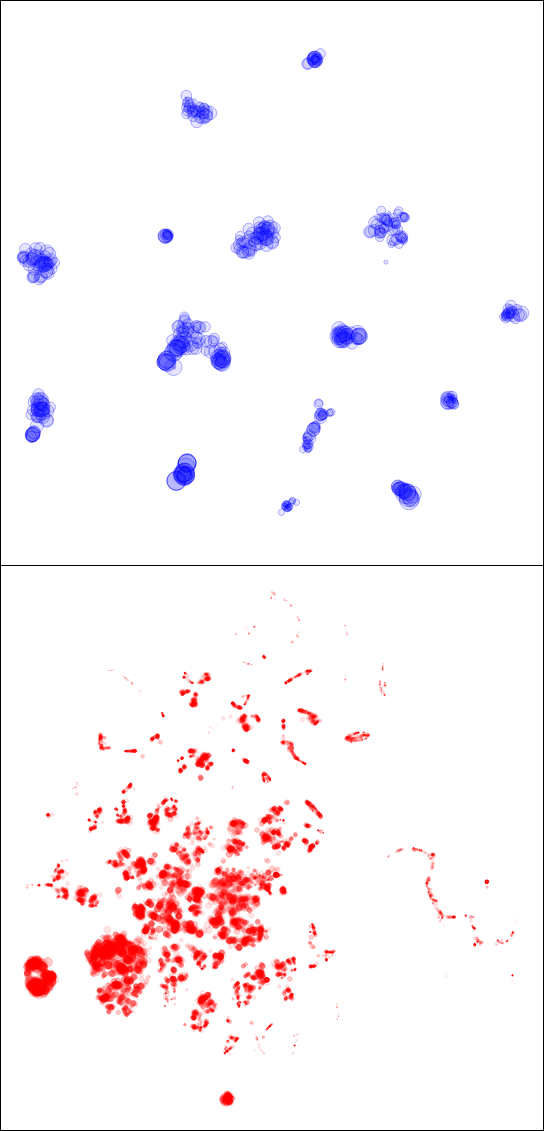}}
    \subfigure[BibUrl]{\includegraphics[width=0.19\textwidth]{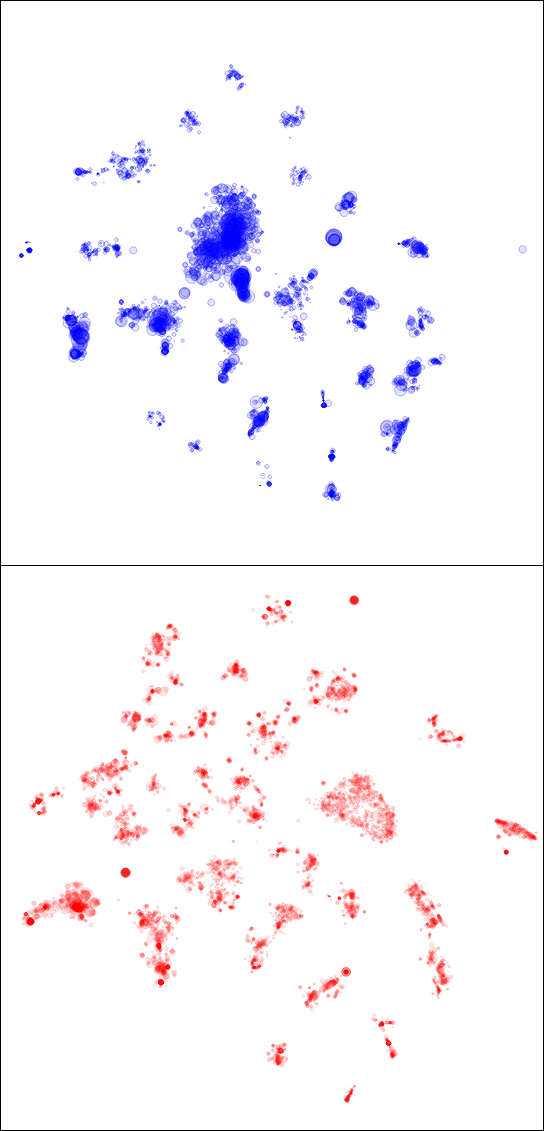}}
    \subfigure[BibTex]{\includegraphics[width=0.19\textwidth]{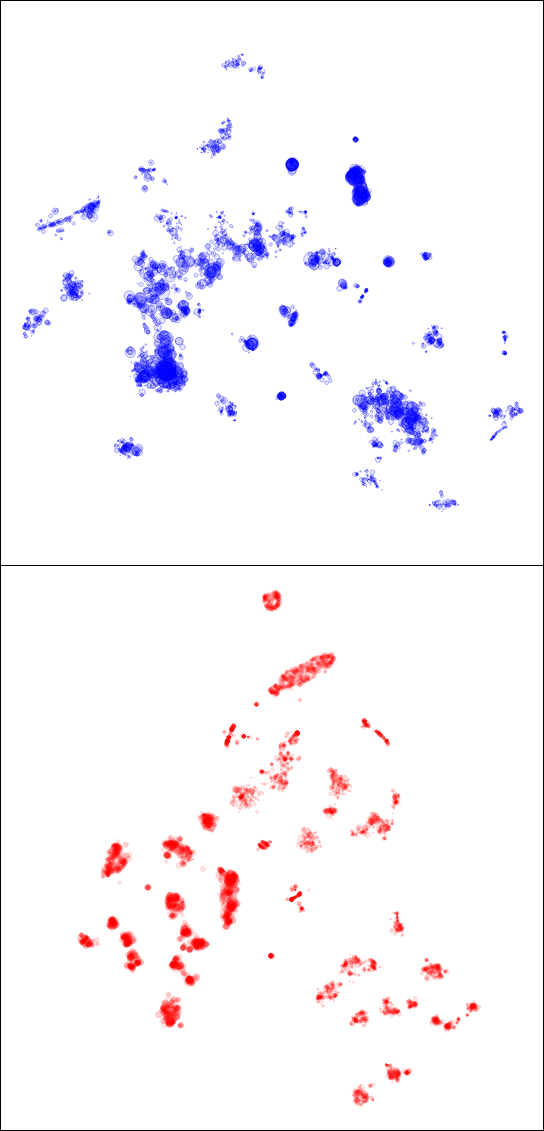}}
    \subfigure[Last.fm]{\includegraphics[width=0.19\textwidth]{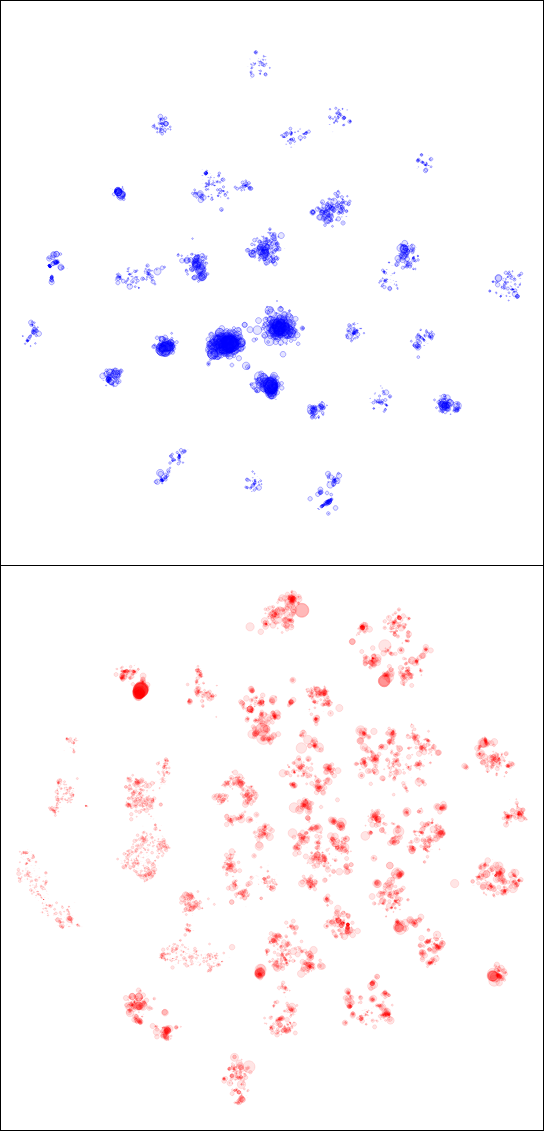}}
    \caption{Co-embedding results.
      The blue items on the top correspond to tag embeddings while the red ones to resource embeddings. Opacity depends on area crowdedness and items' frequency (light color for rare items).
    }
    \label{fig:images_2}
\end{figure}

Fig. \ref{fig:images_2} shows different co-embedding results for the different datasets, where samples' characteristics are presented in Table \ref{tab:CLS}.
The majority presents disjoint clusters for both tags and resources.
The tag embeddings differ from the resource embeddings as they all show a large central cluster.
These clusters correspond to unspecific vocabulary, which occurs in all resources' clusters.
There is no such a central cluster for resources' clusters, unless for the unfiltered \textit{DBLP} subset and \textit{Delicious}.

Datasets are filtered to fit in memory by discarding rare items as they are the less accurate and would lead to a \textit{normalization bias} (\ref{sec:normbias}).
We illustrate the difference between the \textit{unfiltered} (u) and \textit{filtered} (f) over a  \textit{DBLP} subset related to the \textit{payment}.
The filtering leads to removing many infrequent tags,  drastically reducing $|\mathcal{T}|$  and $|\mathcal{C}_T|$.
$|\mathcal{R}|$ is unaffected by the filtering, but $|\mathcal{C}_R|$ is reduced by $1/7$, leading to denser clusters.
The reduction of $|\mathcal{C}_R|$ results from removing these infrequent tags that isolated some resources into small clusters.

Some datasets have a different clusters' shape and spatial organization.
At the same time, most tag clusters are dense; the \textit{Corel5K} has very thin clusters.
This could be explained by the very low $\mathbb{E}(|r|)$, and the small tag collection, leading to very weak connectivity.
The \textit{Bibsonomy} datasets have some of their tag clusters connected to each other.
These clusters gather similar vocabulary but with different formatting.
By applying text processing methods (steaming and case normalization), these clusters would merge in one.

\subsection{Cluster Balance}

The characteristics of the selected subsets used to build the co-embedding presented in Fig. \ref{fig:images_2} are detailed in Table \ref{tab:CLS}.
The subsets dimensions $|\mathcal{R}|$ and $|\mathcal{R}|$ are limited to thousands of items to fit in memory.
The table gathers the number of clusters ($|\mathcal{C}_R|$ an $|\mathcal{C}_T|$) obtained by clustering the co-embeddings ($CE$) with Mean-Shift, and the effective number of clusters $k(\mathcal{C}^{CE})$ measured with Eq. \eqref{eq:coemb_entropy}.
The MS clustering results are compared to the raw binary matrix's spectral clustering ($SC$), searching for the same number of clusters $|\mathcal{C}|$.

\begin{table}[hbt!]
  \caption{Sample size, number of clusters, and effective number of clusters using our co-embedding approach $(CE)$ and spectral clustering $(SC)$. }
  \label{tab:CLS}
  \centering
  \begin{tabular}{|l||cc|ccc|ccc|}
    \hline
    Dataset & $|\mathcal{R}|$ & $|\mathcal{T}|$ & $|\mathcal{C}_R|$ & $k\left(\mathcal{C}_R^{CE}\right)$ & $k\left(\mathcal{C}_R^{SC}\right)$ & $|\mathcal{C}_T|$ & $k\left(\mathcal{C}_T^{CE}\right)$ &  $k\left(\mathcal{C}_T^{SC}\right)$ \\
    \hline
    BibTex    & 7,643     & 4,335 & 54   & \textbf{35.4} & 20.7    & 45  & \textbf{35.9}    & 15.9  \\
    DBLP (u)  & 5,028     & 5,624 & 114  & \textbf{71.8} & 48.1    & 102 & \textbf{76.9}    & 41.95  \\ 
    DBLP (f)  & 4,948     & 1,443 & 81  & 40.2 & \textbf{61.4}     & 16  & 13.2    & \textbf{13.5}  \\ 

    Corel5K   & 5,000     & 364   & 88  & \textbf{62.5}  & 60.9    & 14  & \textbf{12.3}    & 7.1  \\
    Flickr    & 7,185     & 2,427 & 70  & \textbf{34.2}  & 26.7    & 16  & \textbf{13.0}    & 1.9 \\

    MovieLens & 10,159    & 1,127 & 21  & 9.5  & \textbf{18.9}     & 10   & 7.8    & 7.8 \\
    IMDB      & 16,000    & 1,001 & 125 & 65.3 & \textbf{90.1}     & 8    & \textbf{6.4}    & 5.7 \\

    Delicious & 16,105    & 501   & 118  & \textbf{76.2} & 41.1    & 9    & \textbf{8.4}    & 4.2  \\
    BibUrl    & 7,886     & 3,648 & 45   & \textbf{25.3} & 9.8     & 38   & \textbf{27.6}   & 9.0  \\

    Last.fm    & 9,897    & 2,516 & 66  & \textbf{32.9}  & 15.5    & 32 & \textbf{29.5}    &  5.8 \\

    NG20      & 13,125    & 1,007 & 59  & 27.8  & \textbf{48.5}    & 9   & \textbf{6.8}    & 4.2 \\
    OHSUMED   & 13,929    & 1,002 & 34  & 26.7  & \textbf{31.9}    & 7   & 5.1    & \textbf{6.4} \\

  \hline
\end{tabular}
\end{table}

As a general remark, tag clusters are less numerous than resource clusters as observed in the previous figure.
This could be explained by the fact that $|\mathcal{R}| > |\mathcal{T}|$ for all subsets.
However, there is no explicit relationship between sample size and clusters number, as \textit{DBLP} has more resources clusters than most  other datasets while having the smallest $|\mathcal{R}|$.
Thus, the number of clusters found is likely to depend on the sample intrinsic characteristics.

$k\left(\mathcal{C}_T\right)$ is almost always larger for $CE$ than for $SC$, while half of the time for $k\left(\mathcal{C}_R\right)$.
$k(\mathcal{C}^{CE})$ and $k(\mathcal{C}^{SC})$ have the same order of magnitude, unless for a few examples like \textit{Flickr} and \textit{Last.fm} with a very low $k(\mathcal{C}_T^{SC})$ compared to $|\mathcal{C}_T|$, and \textit{BibUrl} where both $k(\mathcal{C}_R^{SC})$ and $k(\mathcal{C}_T^{SC})$ are much lower than for $CE$.

One reason that might explain the difficulty to identify balanced clusters using the raw binary matrix with $SC$ is the power-law distribution of items.
This distribution leads to a very sparse matrix where groups are difficult to identify.
Resource occurrences follow a power-law if some resources are more popular therefore more tagged than others, but the phenomenon is more frequent for tags because of the language organization. This might explain why $SC$ gives lower results than $CE$ more frequently over tags than over resources.

\subsection{Representative Keywords}

The top 4 words of each cluster were extracted using Eq. \ref{eq:representativeness} from the clusters identified previously using $CE$.
As a result, the keywords of the largest clusters (denoted \textit{Main group}) and four other selected clusters are listed in Table \ref{tab:tag_cluster}.

\paragraph*{Main Group Description}

The main cluster of each subset is mostly made of general words occurring in many different contexts that the dataset cover.
In some datasets, it allows to define the global scope (music genres are identified for \textit{Last.fm}, medical topics for \textit{OHSUMED}, or web components for \textit{BibUrl}), but for the majority the media type (music, images, etc.) or the topics covered are difficult to identify.

\paragraph*{Cluster Identity}

Each cluster has a particular identity.
For example, for \textit{OHSUMED}, one group gathers the vocabulary used in studies (where cohorts are compared), and the others concern different disciplines (related to heart, infectiology, and surgery).
For \textit{Corel5K}, each group corresponds to a particular picture (seaside, tundra, locomotion engines, and animals).

A folksonomy includes vocabulary from users that do not necessarily share the same language.
\textit{Flickr} is the most representative dataset, with French, Italian, English and German words each gathered in a specific cluster.
This particularity is visible on the other folksonomies \textit{BibTex}, \textit{BibUrl} and \textit{Delicious} with English and German clusters.

In a given dataset, words groups can be of various kinds, out of language consideration.
For example, the \textit{Last.fm} dataset has five clusters with distinct words types and ideas.
The main group is about general  music type while the 4th gathers sub-music types related to electronic dance music, the $2^{nd}$ gathers band names, the $3^{rd}$ epochs, and the last road-trip related ideas.

\paragraph*{Data Collection Specificity}

While several datasets have in common the type of tagged resources; the clusters obtained differ in the ideas expressed.
This could be explained in a first place by the small overlap between the dataset's vocabulary, but the ideas expressed differ from one dataset to another.

\textit{Corel5K} gathers timeless tags each representing a specific object or place,
while \textit{Flickr} gathers tags related to art requiring culture to be understood.
The difference can be explained by the different goals expected, where \textit{Corel5K} is designed for Object Recognition, while information retrieval for Flickr.

The largest difference concerns \textit{IMDB} and \textit{MovieLens} that both correspond to movies' descriptions with approximately the same number of tags.
Movie styles are easily identified on the \textit{MovieLens} dataset (horror, science-fiction, historical and police movies),
while \textit{IMDB}' clusters group elements that occur in the same context together.
Each dataset has a narrowed list of terms, corresponding to the most relevant term for \textit{IMDB}, and general movies attribute \cite{movielens_20m} for the \textit{MovieLens} tag genome, explaining the difference of expressed ideas between the two datasets.

\begin{table}[hbt!]
  \caption{Top tags for the largest cluster (denoted \textit{Main group}) and four other clusters for each dataset. Acronyms: CS: Computer Science, CVA: Common Value Auction, HCI: Human Computer Interaction, NN: Neural Networks, Mech.: Mechanism}
  \label{tab:tag_cluster}
  \resizebox{\columnwidth}{!}{
  \begin{tabular}{|l||c|c|c|c|c|}
    \hline
    Dataset & Main group & Group 2 & Group 3 & Group 4 & Group 5 \\
    \hline

    BibTex    & Middle   & Recording   & Cell     & Inflation    & Joint       \\
              & of       & Videotape   & Protein  & Geld         & Hip         \\
              & Humans   & Video       & Models   & Kapitalmarkt & Dislocation \\
              & Cerebral & Equilibrium & Membrane & EU-staaten   & Ostonomy    \\
              \hline

    DBLP (f)  & CS                   & CVA           & Use case     & 3-D Secure         & Multimedia \\
              & Payment              & Revenue                & USable       & Credential         & Analytics \\
              & DB transaction       & Op. research    & Web service  & MULTOS             & HCI     \\
              & The Internet         & Mech. design       &  Web app.    & OpenPGP card       & Artif. NN \\
              \hline

    Corel5K   & water & canal     & tundra & train      & deer          \\
              & sky   & sailboats & polar  & plane      & elk           \\
              & tree  & lake      & bear   & railroad   & white-tailed  \\
              & hills & dock      & storm  & locomotive & antlers       \\
              \hline

    Flickr    & light  & pontrouge       & mare     & meiji  & spiegelung \\
              & blue   & pontneuf        & nuvole   & era    & landskap   \\
              & nikon  & collette        & tramonto & period & flickrolf  \\
              & city   & impressionniste & acque    & prints & reise      \\
              \hline

    MovieLens & talky          & splatter & scifi  & us history      & crime          \\
              & cinematography & gory     & space  & war             & thriller \\
              & criterion      & horror   & sci-fi & wartime         & mystery   \\
              & melancholic    & demons   & sci fi & ethnic conflict & murder   \\
              \hline

    IMDB      & life           & films       & american     & father  & game    \\
              & year           & director    & film         & finds   & team    \\
              & time           & video       & movie        & wife    & star    \\
              & day            & documentary & war          & friends & battle  \\
              \hline

    BibUrl    & web       & Post-abortion & assistive & stanford       & kittens   \\
              & blog      & Pro-life      & impaired  & ml             & cats      \\
              & reference & Lebensrecht   & visually  & bioinformatics & memetics  \\
              & blogs     & Human Rights  & cane      & model          & cute      \\
              \hline

    Delicious & throat  & das & breath & apache  & him        \\
              & knees   & ein & loud   & article & looked  \\
              & leaning & den & mouth  & rails   & told     \\
              & rodney  & der & neck   & ruby    & couldn  \\
              \hline

    Last.fm   & pop              & Hamburger Schule & sixties & tech-house  & country pop     \\
              & indie            & The Killers      & 1960s   & seep techno  & switzeland     \\
              & rock             & lofi             & 1960's  & melodic trance    & discover  \\
              & alternative      & elliott smith    & 70's    &  dream trance  & daytrotter   \\
              \hline

    NG20      &  time     &  work   &  religious & windows   &  law \\
  	          &  don      & problem &  jewish    & software  &  country \\
  	  	      &  make     & mail    &  religion  & file      &  government \\
              &  article  & system  &  god       & ms        &  rights \\
              \hline

    OHSUMED   & patients & compared & artery   & antibody & surgery       \\
              & disease  & greater  & left     & antigen  & surgical      \\
              & findings & 4        & cardiac  & immune    & complication  \\
              & clinical & group    & anterior & antibody & procedure    \\
              \hline

\end{tabular}
}
\end{table}

\clearpage
\pagebreak

\subsection{Cluster Retrieval}

We tested the cluster retrieval tasks using the clusters obtained previously.
In addition, we tried the task in both directions: searching the resource's cluster, and searching the tag's cluster.
Table \ref{tab:MRR} summarizes the retrieval scores for this task.

\begin{table}[hbt!]
  \caption{Mean Retrieval Rank for resources clusters $\mathcal{C}_R$ and tag clusters $\mathcal{C}_T$ using our approach ($CE$) and the comparative algorithm ($SC$).}
  \label{tab:MRR}
  \centering
  \begin{tabular}{|l||cc|cc|}
    \hline
    Dataset  & $MRR(\mathcal{C}_{R}^{CE})$  & $MRR(\mathcal{C}_R^{SC})$  & $MRR(\mathcal{C}_T^{CE})$ &  $MRR(\mathcal{C}_T^{SC})$ \\
    \hline
    Bibtex       & \textbf{94.1} \% & 89.1 \%   & \textbf{92.5}   \% & 89.6 \% \\
    DBLP (u)     & \textbf{86.7} \% & 70.3 \%   & \textbf{97.8}   \% & 73.6 \% \\ 
    DBLP (f)     & \textbf{91.5} \% & 66.2 \%   & \textbf{99.6}   \% & 88.7 \% \\ 

    Corel5K      & \textbf{69.2} \% & 48.5 \%  & \textbf{99.3}   \%  & 90.46 \%   \\
    Flickr       & \textbf{78.5} \% & 20.9 \%  & \textbf{96.2}   \%  & 86.74 \%  \\

    IMDB         & \textbf{96.4} \% & 34.7 \%  & \textbf{100.0} \%   & 92.9 \%  \\
    MovieLens    & 62.8 \% & \textbf{79.2} \%  & 86.9  \%   & \textbf{93.6} \% \\

    BibUrl       & 87.1 \% & \textbf{89.0} \%  &  90.4 \%   & \textbf{91.7} \% \\
    Delicious    & \textbf{74.9} \% & 53.7 \%  &  \textbf{98.1} \%   & 97.5 \% \\

    Last.fm      & \textbf{83.8} \% & 70.0 \%   & \textbf{97.9}  \%  & 85.2 \%  \\

    NG20         & \textbf{88.7} \%  & 49.9 \%  & \textbf{98.2}   \% & 96.1 \% \\
    OHSUMED      & \textbf{91.4} \%  & 50.6 \%  & \textbf{97.9}   \% & 91.5 \% \\

  \hline
\end{tabular}
\end{table}

Tags clusters have higher retrieval scores than resources clusters with both approaches, ours leading to better MRRs in most cases.
The good MRRs  are explained by the low number of tag clusters, making the task easier with a smaller choice.
The other point explaining these MRRs is many resource clusters, allowing a more detailed description vector, leading to a more accurate retrieval.
The argumentation is reversed for resource cluster retrieval, leading to lower MRR scores.
The \textit{Flickr}  dataset mentioned  for its low $k(C_T^{SC})$ ($1.9$) illustrates this argument with the lowest retrieval score $MRR(\mathcal{C}_R^{SC}) = 20.9 \%$.
A low number of tags clusters inevitably leads to a low retrieval rate of the resource clusters.
Still, a large number doesn't guarantee the opposite outcome, illustrated by the \textit{DBLP} dataset, with an acceptable retrieval score using our approach while having the largest number of tag clusters.

The gap between approaches for resource retrieval is larger, with our approach having MRR around $80 \sim 90 \%$, the spectral clustering has MRR around $50 \sim 70 \%$.
These results make our approach preferable for this particular task, with the additional advantage of providing an intermediate visualization result.

\section{Discussion}

\subsection{Parameters Choice}

\paragraph{Perplexity:}

As said previously, the perplexity governs the embedding shape by controlling the number of items $t$-SNE would look at for embedding data.
The values we used need to be relatively small as we are interested in local structures to learn items similarity on it.
These values are adjusted as a function on the sample size $n$ the following way:
$$
perp = \left\{
    \begin{array}{ll}
        15 & \text{if } n < 500  \\
        30 & \text{if }  500 \leq n \leq 8000 \\
        60 & \text{else}
    \end{array}
\right.
$$
This decision rule is very simple and allows a simple parameter selection.
Nevertheless, the verification of the produced output is suggested.
A too large value would tend to connect the different clusters together, while a smaller value would lead to more clusters, which both can penalize the other embedding.

\paragraph{Relationships between $\sigma$ and $perp$:}

The estimation of $\sigma$ using Eq. \eqref{eq:radius} can be explained with two remarks.
First, a larger perplexity leads to smaller distances between nearest neighbors (NN), making the choice of $\sigma$ non-trivial.
The direct adaptation to the embedding allows to adapt to the perplexity and to the dataset characteristics.
Second, the perplexity governs the number of NN that $t$-SNE considers  around each item.
If $k < perp$, the kernel would underexploit the items' positioning, as only a small fraction of the NN would be considered.
On the opposite case where $k > perp$, the kernel would be too large, and would put weights on items that where not considered by $t$-SNE. Additionally, a larger $\sigma$ would lead to a flatter kernel, underexploiting items well positioned.
Therefore, the best option is to select $k = \lfloor perp \rceil$,
simplifying $\sigma$'s choice by adapting it indirectly to the current perplexity.

\subsection{Normalization}

\paragraph{Normalization Bias:} In Eq. \eqref{eq:proba}, the contribution of an item is normalized by its frequency, limiting the contribution of highly popular items.
In contrast, items with very few links and low frequency have their contribution  enhanced.
For very low frequency items, their weight is very large,
leading to \textit{group artefacts}, where the unfrequent item is strong enough to gather co-occurring items in a single cluster.
To overcome this issue, using a different normalization scheme or filtering the dataset could mitigate the apparition of artefacts.

\label{sec:normbias}

\paragraph{Weighted Case:}
Our approach exclusively considers the case where the frequency information is unavailable.
Nonetheless, the Eq. \eqref{eq:proba} can be adapted to consider items' frequency, by replacing $\sum_{f_i \in \mathcal{F}} \frac{\delta(s, f_i)}{|f_i|}$ by $\sum_{f_i \in \mathcal{F}} \frac{TF(s, f_i)}{\sum_{s_j \in \mathcal{S}} TF(s_j, f_i)}$  where $TF(s_j, f_i) = \frac{ C(s_j, f_i) }{ \sum_{f' \in \mathcal{F}} C(s_j, f')}$
is the frequency of feature $f_i$ in sample $s_j$ and $C(s_j, f_i)$ the number of occurrences.
If we have words and documents, the normalization the words count within a document is intuitive, allowing to obtain a document vector summing to $1$.
However, describing a word using the documents that contains it to obtain a word vector is non-trivial as different possibilities exist.
One can normalize the word frequency of each document $\frac{TF(w, r)}{\sum_{r' \in \mathcal{R}} TF(w, r')}$, or one can normalize the word count per document over all the corpus $\frac{C(w, r)}{\sum_{r' \in \mathcal{R}} C(w, r')}$.
The first option is more balanced, as it gives credit to all documents. Nevertheless, for documents with very few words, it would penalize the final result because it considered inaccurate information. In that case, the second option is more suitable, as it favors the contribution of documents of sufficient size.

\subsection{Scalability to Large Datasets}

Each step of our algorithm runs in $\mathcal{O}\left(nm (n+m) + kn^2\right)$, unaffordable for very large datasets.
Nevertheless, the scalability problem can be bypassed by identifying clusters over a subset of items.
This assume that a sampling would only affect clusters' size, but would not affect clusters' number.
After the cluster identification phase, the unselected items are assigned to the most relevant cluster, allowing to classify them.

Then, the embedding process could be repeated over the items of a consolidated cluster, leading to a new segmentation level.
Some features were previously discarded because of their low frequency within the initial subset because of the filtering process.
The addition of these new items allows some of these features to reach the critical frequency, therefore to consider them.
Even if the items are very similar, the second co-embedding process is likely to lead to new clusters because of the new features that increase diversity.

\subsection{Adaptation to Temporal Datasets}

A dataset is a snapshot of a database at a given moment.
A real-world graph is often dynamic, with new nodes and new edges.
Rather than looking at the entire network since origin,
the focus can be put over a given period to compare with another one.

The proposed approach can be adapted to temporal datasets using a moving window.
The first slice is used to construct a primer co-embedding.
Next, the embedding of the new window is obtained using the embeddings of the previous slice to learn items' density.
This step is possible only if the previous and current windows share some of their elements, and that new items have already seen features.
Otherwise, it would not be possible to evaluate the density.
By following this process, it would reduce the number of co-embedding iteration as an already stable embedding is provided to estimate items' density.
A minor difficulty concerns cluster tracking over the different windows, as new clusters can emerge, disappear, split or fuse together, adding some complexity.

\section{Conclusion}

In this article, we proposed a co-clustering algorithm for bipartite graphs.
The algorithm addresses \textit{dimension relatedness} by projecting features into a low dimensional space, comparing samples based on their mixture of features.
The embedding process leads to natural cluster formation for a dataset where community structures exist, clustered  using the Mean-Shift algorithm.
The algorithm is easy to configure with very few parameters to adjust.
We tested our algorithm over a cluster retrieval problem.
A better retrieval accuracy with more balanced clusters in a large number of cases was shown than when using a spectral co-clustering algorithm that did not consider the relationship between the dimensions.
However, the weakest point of our algorithm is scalability.
Nevertheless, clusters could be identified over a subset, and unused items assigned to the closer cluster.
The approach could be used to optimize recommender systems and retrieval engines by working at the cluster-level rather than the item-level.

\bibliographystyle{unsrtnat}
\bibliography{references}

\section*{Appendix}

\label{section:appendix}

\subsection*{Sampling \& Filtering}

Because of the complexity of our proposed method, not all the datasets could be processed as they couldn't fit in the memory (Bibtex, DBLP, Flickr, IMDB, and Last.fm).
Therefore, a first sampling is performed over the documents, trying to keep around 10,000 documents.
Then, some documents having too few keywords (less than $5$) are removed from the subset. The same filtering is applied to keywords with less than $5$ occurrences.
The filtering process is repeated until all documents have $5$ valid tags and reciprocaly.

\subsection*{Dataset Processing}

\paragraph{DBLP}
We used the DBLP version 12 (2020-04-09) citation dataset \cite{DBLP}.
The dataset corresponds to scientific articles with information such as title, date, abstract, authors, references, around 10 keywords and other fields.
The bipartite graph was build using the associated keywords.
Because of the dataset size, a subset of article was extracted.
A general keyword was selected (in the experiment: \textit{Data center}), and all documents associated with this keyword gathered.
Then, the sampling is performed followed by the filtering step.

\paragraph{Bibtex}
The Bibsonomy dataset \cite{bibsonomy} is composed of two sub-dataset: the former correspond to tagged URLs, while the second to tagged scientific articles.
We use the scientific part, denoted Bibtex.
This dataset is a folksonomy, where users are free to add any keywords to any documents.
We construct the bipartite graph by preserving the (document, tag) part, forgetting the user part.
Because of the folksonomy origin, the keywords are power-law distributed.
We couldn't perform an initial document selection looking at a particular keyword, because the resulting subset after filtering would be too small.
Therefore, documents were sampled at random from the whole corpus.

\paragraph{Flickr}
We used the Flickr dataset \cite{huiskes08} (MIRFLICKR-1M) that contains 1 million of images and the corresponding tags.
We removed from the described corpus all images with no associated tags.
Then, we applied the sampling and filtering steps.

\paragraph{Corel5K}

The Corel5K dataset \cite{corel5k} contains 5 thousand images associated with tags.
The dataset has been obtained on the Cometa website (\url{https://cometa.ujaen.es/datasets/}).
Each image is associated with a set of tags, where no frequency is available.
Due to the limited size of this dataset, no sampling nor filtering have been done.

\paragraph{IMDB}

The IMDB dataset \cite{IMDB_cometa} has been gathered on the Cometa website.
Each movie is associated with a set of tags as well as a set of classes, representing the movie types.
The bipartite graph is constructed using the tags and movies exclusively, and does not use the movie types.
For the experiments, the dataset was sampled and filtered.

\paragraph{MovieLens}

From the MovieLens project \cite{movielens_20m}, we selected the  ml-20m dataset and studied the \textit{genome} part.
The \textit{genome} is a set of about 1.000 keywords for which the relevance to each movie ($\in [0, 1]$ has been evaluated.
For each document, we preserved all the keywords with relevance equal or greater than $0.5$.
The subset size has been sampled and filtered for the experiments.

\paragraph{Last.fm}

The Million Song Dataset \cite{million_song_dataset} corresponds to music titles associated with rated attributes in $[0, 100]$.
The dataset is organized into three part: \textit{train}, \textit{test} and \textit{subset}.
We worked only on the \textit{train} part because of the large number of music songs available.
As for the MovieLens dataset, we preserved all the attributes with relevance $\geq 80$, and discarded the songs with no items.
This dataset has been sampled and filtered for the experiments.

\paragraph{NewsGroup20}

The NewsGroup20 dataset \cite{newsgroup_cometa} has been obtained on the Cometa website.
For scalability, we only removed the documents with the lowest number of tags.
We did not apply the filtering process as the number of tags per document is large, as well as the number of documents including a specific tag. Therefore, the filtering would have no consequence.

\paragraph{OHSUMED}

The OHSUMED dataset \cite{ohsumed_cometa} corresponds to medical article abstracts.
It has been obtained on the Cometa website without processing.

\end{document}